\documentclass{elsart}
\input epsf
\newcommand{\beq}{\begin{equation}}
\newcommand{\eeq}{\end{equation}}
\newcommand{\br}{\hfil\break}
\newcommand{\ep}{\epsilon}
\newcommand{\la}{\lambda}
\newcommand{\mod}{\,\mbox{mod}\,}
\def\prt #1#2{{\partial #1 \over \partial #2}}

\begin{document}
\begin{frontmatter}

\title{Hexagonal patterns in finite domains}

\author{P.C. Matthews}
\address {Department of Theoretical Mechanics, University of
Nottingham,  Nottingham NG7 2RD,  United Kingdom}

\begin{abstract}
In many  mathematical models for pattern formation, a regular
hexagonal pattern is stable in an infinite region.
However, laboratory and numerical experiments are carried out 
in finite  domains, and this imposes certain constraints on the
possible patterns. 
In finite rectangular domains, it is shown that a regular
hexagonal pattern  cannot occur if the  aspect ratio is rational.

In practice, it is found experimentally that in a rectangular region, 
patterns of irregular hexagons are often observed. 
This work analyses the geometry and dynamics of irregular hexagonal
patterns. These patterns occur in two different symmetry types, either
with a reflection symmetry, involving two wavenumbers, or without
symmetry, involving three different wavenumbers.

The relevant amplitude equations are studied to investigate the
detailed bifurcation structure in each case.
It is shown that hexagonal patterns can bifurcate subcritically either from
the trivial solution or from a pattern of rolls.

Numerical simulations of a model partial differential equation are
also presented to illustrate the behaviour.

\end{abstract}


\end{frontmatter}

\section{Introduction}

Pattern formation is a topic of intensive current research; a recent
volume of this journal has been dedicated to the subject \cite{Physd}.
There are many experimental examples of pattern formation, including
Rayleigh--Benard convection \cite{Gol84}, directional solidification
\cite{Hoyle96} and the Faraday experiment \cite{Kud96}.
It is now appreciated that the amplitude equations describing the
evolution and dynamics of patterns are generic, so that many different
physical systems are governed by the same equations. These amplitude
equations and their possible solutions are determined more by the
symmetries of the  problem than the details of the physics
\cite{CrawK91}. 

Theoretical work on pattern formation has generally been concerned 
with regions of infinite horizontal extent, with attention focussed on
modes with a single wavenumber. 
Analysis of the competition between rolls and hexagons shows that 
in systems without an up--down symmetry, 
hexagons are generally preferred. For hexagons, a sign change in the
pattern gives a genuinely different solution, while for rolls a sign
change is equivalent to a spatial translation of half a wavelength. 
This means that
hexagons can occur at a transcritical bifurcation but rolls can only
appear at a pitchfork bifurcation.
In systems with an up--down symmetry, only pitchfork bifurcations
occur, and either rolls, squares, or
hexagons may be stable, depending on the coefficients in the amplitude
equations.

Recently, more complicated patterns involving four sets of rolls  with
the same wavenumber have been considered \cite{Dion96}. These share
the symmetry properties of rolls and squares, so a sign change is
equivalent to a spatial shift. 

Analysis of patterns involving a single wavenumber is not always
successful in predicting patterns. Resonant mode interactions can be
important near certain points in parameter space. These resonances
generate additional quadratic terms in the amplitude equations and can
lead to more exotic patterns with complicated dynamics \cite{Proc96}.

In most examples of experimental interest, there is some up-down
asymmetry which leads to a preference for hexagons at the onset of
instability. However, the finite geometry  of the experimental system
or numerical simulation 
does not in general allow a regular hexagonal pattern.
What is observed in practice is a solution which is close to regular
hexagons, but in fact involves  slightly different wavenumbers.
An example is the numerical experiments on compressible
magnetoconvection by Weiss et al. (1996) \cite{Weiss96}, 
who found  patterns of non-regular hexagons in square domains.
Laboratory experiments on surface-tension-driven convection in square 
container  also yield a variety of puzzling patterns    \cite{Kosch90}.

Such irregular or non-equilateral hexagons can also occur in a region
of infinite horizontal extent with anisotropy, since breaking
rotational invariance also breaks the symmetry of the hexagons.
An example of this is convection influenced by a weak shear flow
\cite{Hall95,Cox97}.  Another application is in the numerical modelling
of hexagonal patterns \cite{Skel97}; if the numerical method
does not preserve the symmetry of the hexagons then a qualitatively
incorrect bifurcation diagram can be obtained.

Theoretical work on non-equilateral hexagons has been carried out by 
Malomed, Nepomnyashchy and Nuz \cite{Mal94}, who studied the governing
amplitude equations in two cases. If the pattern involves two equal
wavenumbers, a hexagonal pattern can be stable in a region of
parameter space near the onset of instability, and can branch either
directly from the trivial solution or from a roll pattern. In the case
of three different wavenumbers, hexagons can only be stable at finite
amplitude.
The same amplitude equations have also been studied in the context of
convection in the presence of a mean flow \cite{Hall95,Cox97},
and for an anisotropic solidification problem \cite{Hoyle96}.
However, none of these works has provided complete bifurcation diagrams
showing how the well-known picture for competition between rolls and
regular hexagons is modified by the anisotropy.

This paper describes the geometry of irregular hexagonal patterns
in section 2, considering the constraints imposed by a finite domain
and relating the wavenumbers involved in the pattern to the number of
hexagons seen.
Section 3 considers the nonlinear dynamics of these patterns,
concentrating on those aspects of the problem not covered by previous
work. To illustrate the possible types of behaviour and to provide a
check on the analytical work, numerical
simulations of the modified Swift--Hohenberg model \cite{Swift77} are
described in section 4.

\section{Geometry of hexagonal patterns}
\label{sec:geom}

This section describes the geometry of hexagonal patterns that can be
obtained in a square box with sides of length $L$ with either periodic
or Neumann boundary conditions.

For periodic boundary conditions, any pattern $w(x,y)$ can be written as a
Fourier series,
\beq w = \sum_{m=1}^\infty \sum_{n=1}^\infty  
  A_{mn} \exp 2\pi i( m x + n y) / L  \label{eq:pbc}\eeq
and for Neumann boundary conditions ($\prt w x = 0$ at $x=0$, $L$; 
$\prt w y = 0$ at $y=0$, $L$)  
\beq w = \sum_{m=1}^\infty  \sum_{n=1}^\infty   
  A_{mn} (\cos \pi ( m x + n y)/L + \cos \pi ( m x - n y) / L) . \eeq 
Note that the possible patterns for Neumann boundary conditions are a
subset of those for  periodic boundary conditions in a box twice as
large. This fact is often referred to as a `hidden symmetry' 
\cite{Craw91,Craw93}.
Henceforth  periodic  boundary conditions will be assumed, since this 
includes the other case. The integers $m$ and $n$ in (\ref{eq:pbc})
will be referred to as `wave integers'. The corresponding
wavenumber is 
\beq k = 2\pi \sqrt{m^2+n^2} / L  . \label{eq:k}\eeq

A hexagonal pattern is obtained by taking three modes whose wave
integers sum to zero, i.e.
\beq m_1 + m_2 + m_3 = 0,  \qquad n_1 + n_2 + n_3 = 0 . \label{eq:sumzero}\eeq
This is necessary if `strong' resonance (i.e.\ a quadratic term) 
is to occur in the amplitude equations.

A pattern of regular hexagons with three equal
wavenumbers cannot occur in a square box.
This result is not immediately obvious (in general the hexagons could
be aligned at any angle to the box) but can be demonstrated as
follows.  If the three wavenumbers are equal then
\beq
m_1^2+n_1^2 = m_2^2+n_2^2 = m_3^2+n_3^2 , \label{eq:keq}
\eeq
which can be rewritten using (\ref{eq:sumzero}) as 
\beq
m_1^2+n_1^2 = m_2^2+n_2^2 = (m_1+m_2)^2+(n_1+n_2)^2 .
\eeq
Expanding the brackets, these equations simplify to
\beq
m_1^2+n_1^2 = m_2^2+n_2^2 = -2 (m_1 m_2 + n_1 n_2),
\eeq
showing that the sums of the squares of the wave integers are even.
However it can be stipulated that not all the wave integers are even, 
since any common factors can be removed. 
There are two possibilities, each of which lead to a contradiction.
If $m_1$, $n_1$ are odd and $m_2$, $n_2$ are even, 
then $m_1^2+n_1^2 = 2 \mod 4$ but $m_2^2+n_2^2 = 0 \mod 4$.
Alternatively if $m_1$, $n_1$, $m_2$, $n_2$ are all odd,  then 
$m_1^2+n_1^2 = 2 \mod 4$ but $2 (m_1 m_2 + n_1 n_2) = 0 \mod 4$.

This result, that regular hexagons cannot occur in a square box, 
generalizes to the case of a rectangular box provided that 
the aspect ratio of the box is rational. In this case, if the box is
of size $L_1$ in the $x$-direction and $L_2$ in the $y$-direction
and $L_1 /L_2 = p/q$ where $p$ and $q$ are integers, the constraint of
equal wavenumbers requires
\beq
m_1^2 / L_1^2+n_1^2 / L_2^2 = m_2^2 / L_1^2 + n_2^2 / L_2^2 
                          = m_3^2 / L_1^2 + n_3^2  / L_2^2  . 
\eeq
Multiplying through by $L_1  L_2  \, p\, q$, this becomes
\beq
(m_1 q)^2 + (n_1 p)^2 = (m_2 q)^2 + (n_2 p)^2 = (m_3 q)^2 + (n_3 p)^2
\eeq
which is equivalent to (\ref{eq:keq}) with $m_i$ replaced by 
$m_i q$ and $n_i$ replaced by $m_i p$, so that the problem has been
reduced to the previous case.

Regular hexagons can however be obtained in a rectangular box if
the ratio of the length of one side to the other is a multiple of
$\sqrt{3}$; this choice has been used in numerical simulations to
investigate the relative stability of rolls and hexagons \cite{Matt95}.

\begin{figure}
\centerline{
\epsfxsize0.5\hsize\epsffile{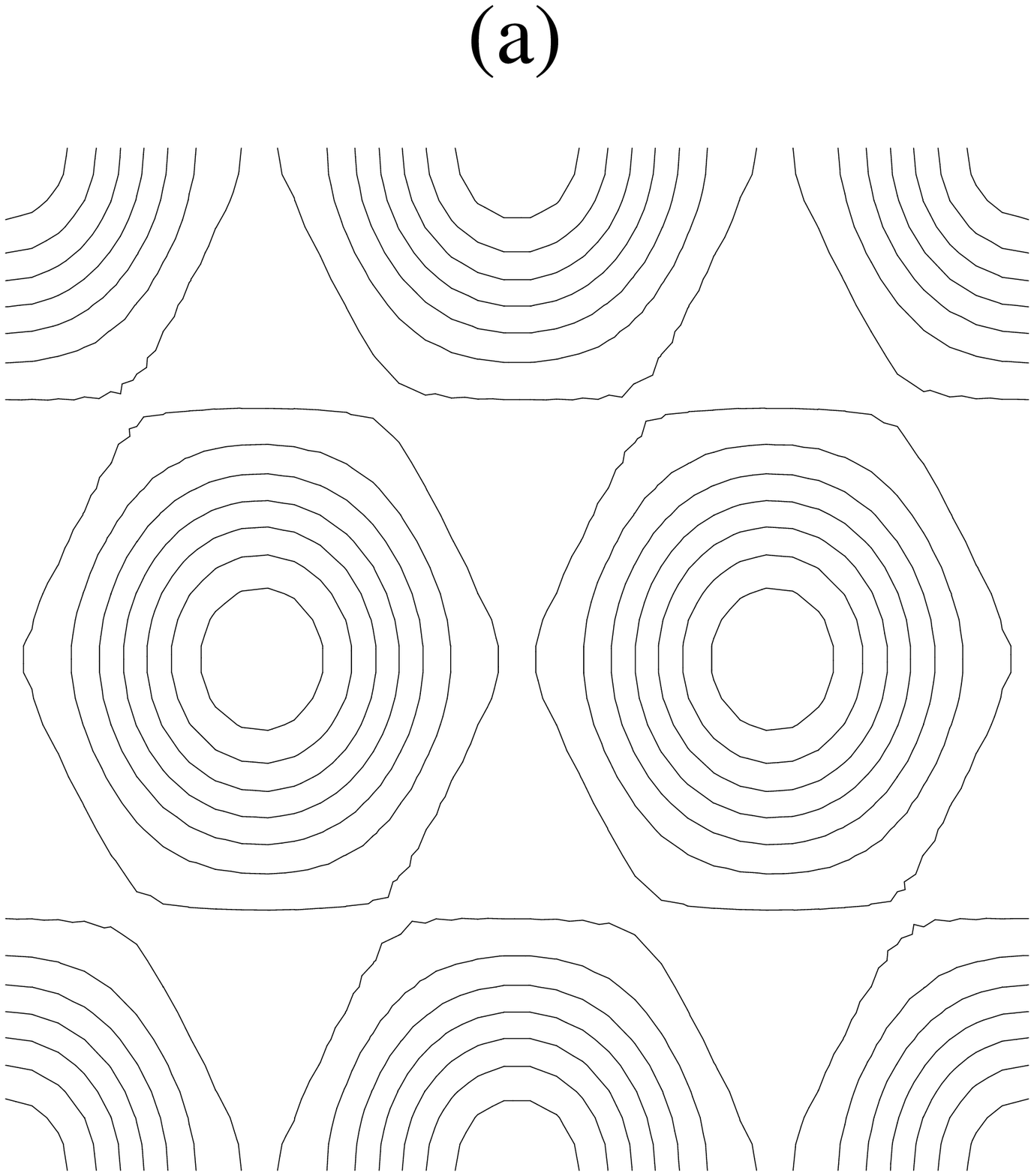}\epsfxsize0.5\hsize\epsffile{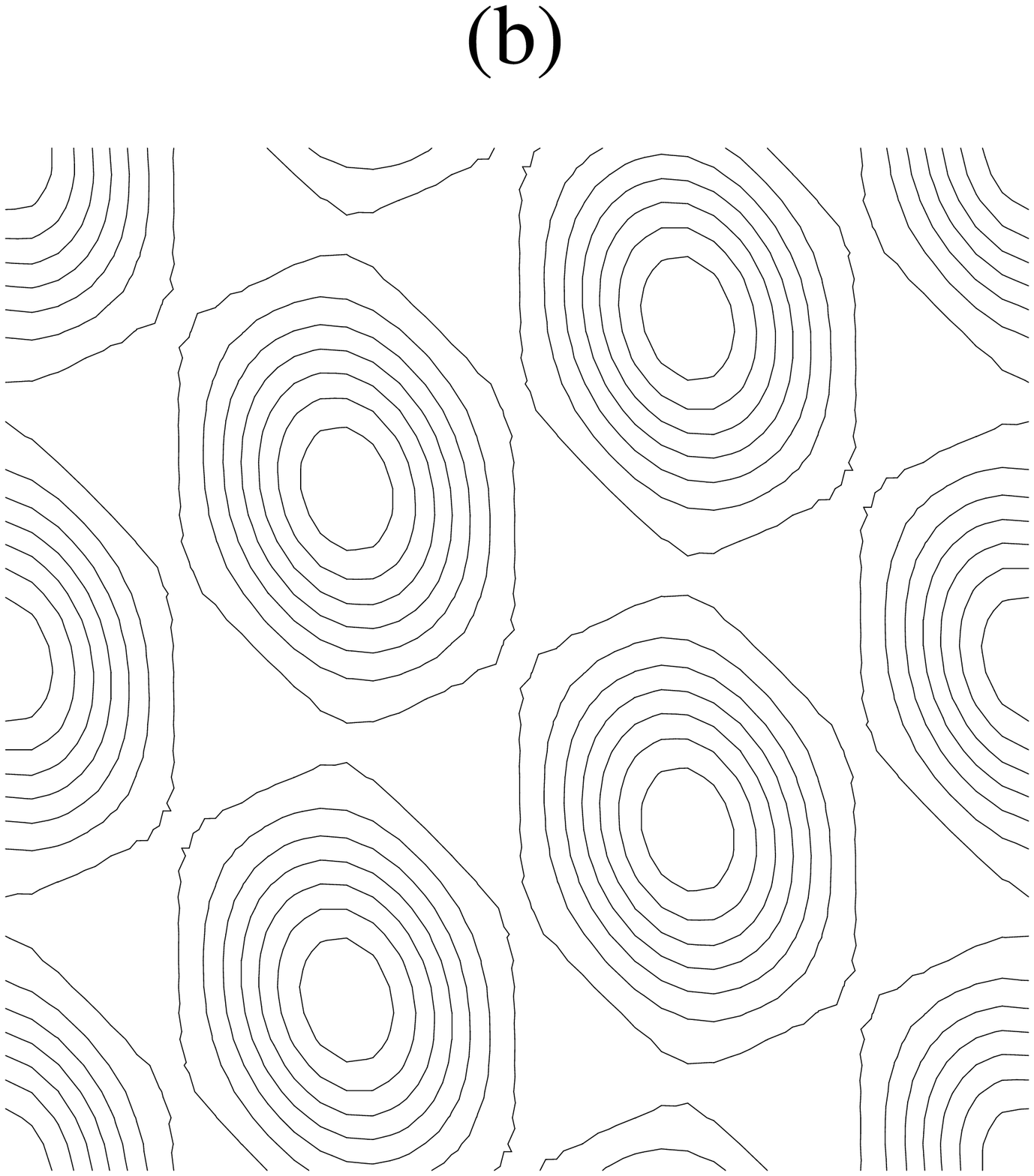}
}
\caption{Hexagonal patterns with wave integers (a) $(2,1)$, $(-2,1)$,
$(0,-2)$; (b) $(3,0)$, $(-2,-2)$, $(-1,2)$.\label{fig:2hex}}
\end{figure}

Two examples of irregular hexagonal patterns are illustrated in 
Fig.~\ref{fig:2hex}. In each case the amplitudes of the three modes have
been chosen to be real and equal, so that the function plotted is
\beq 
w = \sum_{i=1}^3  \cos(2\pi(m_i x + n_i y)/L)  .  \label{eq:3mode}
\eeq
Fig.~\ref{fig:2hex} (a) shows the case
where the wave integer pairs are $(2,1)$, $(-2,1)$ and $(0,-2)$, in which
four hexagons occur in the box, while  the six-hexagon case with 
wave integers $(3,0)$, $(-2,-2)$, $(-1,2)$ is shown in Fig.~\ref{fig:2hex}
(b). The former can occur with periodic or Neumann boundary conditions,
but the latter can only occur with periodic boundary conditions.
The six-hexagon pattern shown in Fig.~\ref{fig:2hex} (b) 
has been found in  numerical simulations of magnetoconvection in a
compressible fluid by Weiss {\it et al.} (1996) \cite{Weiss96}.

It is apparent from these  examples that there are two different
possible symmetries for these  hexagonal patterns. 
The four-hexagon example, with two equal wavenumbers,  has a
reflection symmetry  ($x \to -x$). The six-hexagon case, with three
different wavenumbers, does not have a reflection symmetry. 
It is well known that symmetry has a crucial effect on the nonlinear
dynamics of a system (e.g.\ Crawford  and Knobloch 1991
\cite{CrawK91}), and it will be shown in section \ref{sec:dyn} that 
the bifurcation structure associated with the two patterns is
different. 

There is a simple relationship between the number of hexagons $N$ in the
periodic box and the wave integers. This is 
\beq  N = | m_1 n_2 - m_2 n_1 | .  \label{eq:nhex} \eeq
From (\ref{eq:sumzero}) it follows that any pair of the wave integers
can be used in this formula. 
This result is demonstrated as follows:
the planform function (\ref{eq:3mode}) is maximized when $x$ and $y$
obey
\beq
m_1 x + n_1 y =  p  L , \qquad m_2 x + n_2 y =  q  L,
\eeq
where $p$ and $q$ are integers. From (\ref{eq:sumzero}) it follows
that $m_3 x + n_3 y$ is then also an integer multiple of $L$, 
so that $w$ takes its maximum value of 3, corresponding  
to the centre of a hexagon. These equations give the lattice of points
at which centres of hexagons occur. 
Three points on this lattice are \br
(a) $p=0$,  $q=0$, $x=0$, $y=0$; \br
(b) $p=1$, $q=0$, $x= n_2  L / (m_1 n_2 - m_2 n_1)$, 
$y=- m_2  L / (m_1 n_2 - m_2 n_1)$ ; \br
(c) $p=0$, $q=1$, $x= -  n_1  L / (m_1 n_2 - m_2 n_1)$, 
$y=  m_1  L / (m_1 n_2 - m_2 n_1)$ . \br
These three points form a triangle that connects the centres of three
hexagons. The area of this triangle is
\beq
 | (n_2, -m_2) \times (-n_1, m_1) | \, L^2 / 2  (m_1 n_2 - m_2 n_1)^2 
 =   L^2 / 2 | m_1 n_2 - m_2 n_1 |, 
\eeq
using the formula for the area of a triangle in terms of the cross
product of two vectors. This area is half the area associated with
each hexagon, since each triangle connects three hexagons but each
hexagon is connected to six triangles. 
Dividing the total area of the square $L^2$ by the area of each
hexagon  gives the formula (\ref{eq:nhex}) for the total number of
hexagons in the box. 

\begin{table}
\begin{tabular}{rrrrrrrrr}
 $m_1$& $n_1$ &\qquad $m_2$& $n_2$ & \qquad$m_3$ & $n_3$& 
\qquad\qquad$H$ &\qquad $N$ & Symmetry \\
\noalign{\smallskip\hrule\smallskip}
   2&   1 & -2 &  1 &  0 & -2 & 0.20000 &  4 & Y\\
   3&   0 & -2 &  2 & -1 & -2 & 0.44444 &  6 & N\\
   3&   1 & -1 & -3 & -2 &  2 & 0.20000 &  8 & Y\\
   3&   2 & -3 &  1 &  0 & -3 & 0.30769 &  9 & N\\
   4&   1 & -3 &  2 & -1 & -3 & 0.41176 & 11 & N\\
   3&   3 & -4 &  0 &  1 & -3 & 0.44444 & 12 & N\\
   4&   0 & -2 & -3 & -2 &  3 & 0.18750 & 12 & Y\\
   4&   2 & -1 & -4 & -3 &  2 & 0.35000 & 14 & N\\
   3&   3 & -4 &  1 &  1 & -4 & 0.05556 & 15 & Y\\
   4&   2 & -4 &  2 &  0 & -4 & 0.20000 & 16 & Y\\
   4&   3 & -4 &  1 &  0 & -4 & 0.36000 & 16 & N\\
   5&   1 & -2 & -4 & -3 &  3 & 0.30769 & 18 & N\\
   5&   2 & -4 &  2 & -1 & -4 & 0.41379 & 18 & N\\
   5&   1 & -4 &  3 & -1 & -4 & 0.34615 & 19 & N\\
   5&   0 & -3 & -4 & -2 &  4 & 0.20000 & 20 & Y\\
   5&   2 & -5 &  2 &  0 & -4 & 0.44828 & 20 & Y\\
   5&   2 & -2 & -5 & -3 &  3 & 0.37931 & 21 & Y\\
   5&   3 & -1 & -5 & -4 &  2 & 0.41176 & 22 & N\\
   5&   2 & -1 & -5 & -4 &  3 & 0.13793 & 23 & N\\
   4&   4 & -5 &  1 &  1 & -5 & 0.18750 & 24 & Y\\
   6&   0 & -3 & -4 & -3 &  4 & 0.30556 & 24 & Y\\
   6&   0 & -4 & -4 & -2 &  4 & 0.44444 & 24 & N\\
\end{tabular}
\caption{Wave integer combinations that generate hexagonal
patterns. The parameter $H$ measures the departure from regular
hexagons and $N$ is the number of hexagons in the periodic box. 
The final column indicates whether or not the pattern has a reflection
symmetry.}
\label{tab:hex}
\medskip\hrule 
\end{table}

To describe these hexagonal patterns it useful to have a parameter
measuring how close the pattern is to regular hexagons.
One such parameter is
$ H  =  ( k_{\rm max}^2 - k_{\rm min}^2)/k_{\rm max}^2  $
where $k_{\rm max}$ and $k_{\rm min}$ are the maximum and minimum of
the three wavenumbers; if $H$ is small, the pattern is close to
regular hexagons. 
Table \ref{tab:hex} lists all hexagonal patterns in which 
$H < 0.45$ and $N < 25$, giving the wave integers, the parameter $H$,
the number of hexagons $N$ and the symmetry for each.

\section{Nonlinear dynamics of hexagonal patterns}   
\label{sec:dyn}

This section describes the behaviour of the nonlinear amplitude
equations  governing the dynamics of irregular hexagonal patterns.

Consider a pattern-forming partial differential equation with a
control parameter $r$ and a zero solution that exists for all $r$. 
Suppose that there is a marginal stability curve $r_c(k)$, 
on which the growth rate of perturbations to the zero solution
vanishes, with a single minimum.
There are many examples of such
systems, including convection and the Swift--Hohenberg model
\cite{Swift77}. Let $r_0$ be the minimum value of $r_c(k)$,
and let $k_0$ be the corresponding wavenumber. 
Now for a finite box size $L$, only discrete wavenumbers given by 
(\ref{eq:k}) can occur, and in general these will not include $k_0$.
For large values of $L$ however, wavenumbers close to  $k_0$ will be
included, and the separation between wavenumbers in the vicinity of
$k_0$ scales as $1/L^2$. This allows a small
parameter $\ep$ to be introduced, representing the difference between
$k_0$ and the three wavenumbers that generate the hexagonal
pattern. With the scaling  
$k_i = k_0 + O(\ep)$ for $i=1, 2, 3$, the corresponding values of
$r_c$ are $r_{ci} = r_o + O(\ep^2)$, since $r_0$ is the minimum of $r_c(k)$.
Writing $r = r_0 + \ep^2 r_2$, all 
growth rates are $O(\ep^2)$ so the appropriate
time scale for the amplitude equations is $T=\ep^2 t$.


It is now possible to write down the governing amplitude equations 
for the three Fourier modes obeying the resonance condition
(\ref{eq:sumzero}). Writing 
\beq w = \sum_1^3 A_j \exp(2\pi i( m_j x + n_j y) / L),
\label{eq:expand}
\eeq
it can be shown that the sum of the phases of the $A_j$ tends to zero
\cite{Mal94}, so that with an appropriate choice of origin  the amplitudes 
$A_j$ can be taken to be  real. The scaled amplitude equations are then 
\begin{eqnarray}
\dot A_1 &  = & \la_1 A_1 + A_2 A_3 , \nonumber \\
\dot A_2 &  = & \la_2 A_2 + A_1 A_3 ,  \label{eq:amp} \\
\dot A_3 &  = & \la_3 A_3 + A_1 A_2  .\nonumber   
\end{eqnarray}
Here, the dot represents the rate of change on the slow timescale $T$,
the linear coefficients $\la_1$, $\la_2$ and $\la_3$ are proportional
to  $(r - r_{cj})/\ep^2$ and the amplitudes $A_i$ have been scaled by
a factor $\ep^2$.
With these scalings, all terms in (\ref{eq:amp}) 
are of the same order, and any cubic terms in the amplitude equations
will be smaller by a factor $\ep^2$. 

Note that the coefficients of the quadratic terms in (\ref{eq:amp}) are
equal. This is because (\ref{eq:amp}) represents a small perturbation
from the case of regular hexagons. The coefficients of the quadratic
terms can be set to 1 by scaling the amplitudes $A_j$ appropriately.

The dynamics of the system (\ref{eq:amp}) depends crucially on whether
there are any additional symmetries.
It is useful to begin by briefly reviewing the well-known case of
regular hexagons, for which $\la_1 = \la_2 =\la_3$.
In this case the fixed points of (\ref{eq:amp}) are the solution
$A_1 = A_2 = A_3 = 0$, which is stable for $\la_1 < 0 $ and unstable for
$\la_1 > 0$, and four equivalent hexagonal solutions 
$A_1 = \pm \la_1$, $A_2 = \pm \la_1$, $A_3 = -A_1 A_2 /\la_1$, 
for which the three eigenvalues are $2\la_1$,  $2\la_1$ and $-\la_1$.
The bifurcation at $\la_1 = 0$ is transcritical, 
and the hexagonal solution is never stable. 

In the case where $\la_1 =\la_2$,  two of the
wavenumbers are equal in magnitude and the resultant pattern has a
reflection symmetry (e.g.\ Fig.~\ref{fig:2hex}a).
The fixed points are \\
(i) $A_1 = A_2 = A_3 = 0$. This is stable if $\la_1 <0$ and $\la_3 < 0$, 
and undergoes stationary bifurcations at $\la_1 = 0$ and $\la_3 =  0$. \\
(ii) $A_1 = \pm \sqrt {\la_1\la_3}$, $A_2 = \pm\sqrt {\la_1\la_3} $, 
$A_3 = - A_1 A_2 /\la_3$. 
These four equivalent solutions only exist when $\la_1\la_3 > 0$.
Its eigenvalues $s$ obey $s = 2\la_1 $ or $ s^2 - \la_3 s - 2 \la_1\la_3 =0$.
The product of roots of this quadratic is negative so this solution
can never be stable.  

For the case where all three linear terms in (\ref{eq:amp}) are
different, the three wavenumbers are different and the hexagonal
pattern does not have mirror symmetry (e.g.\ Fig.~\ref{fig:2hex}b).
The fixed points are \\
(i) $A_1 = A_2 = A_3 = 0$. This is stable if $\la_1 <0$, $\la_2<0$ and
$\la_3 < 0$, with stationary bifurcations at $\la_1 = 0$, $\la_2 = 0$ and
$\la_3 = 0$,   \\
(ii) $A_1 = \pm \sqrt {\la_2\la_3}$, $A_2 = \pm\sqrt {\la_1\la_3} $, 
$A_3 = - A_1 A_2 /\la_3$. 
Again there are four solutions of this type.
This solution only exists when either $\la_1 <0$, $\la_2<0$,
$\la_3 < 0$ or $\la_1 > 0$, $\la_2> 0$, $\la_3 > 0$. 
The  eigenvalues $s$ obey $s^3 - (\la_1+\la_2+\la_3) s^2 + 4 \la_1\la_2\la_3 = 0$.
Since there is no linear term, the three eigenvalues obey 
$s_1 s_2 + s_2 s_3 + s_3 s_1 =0$ which shows that the solution can
never be stable. 

\subsection{Amplitude equations including cubic terms}
\label{sec:cubic}

The analysis of the previous section, although asymptotically correct,
has two drawbacks. Firstly, no stable nonlinear solutions are found,
and secondly, solutions in  the form of rolls (involving a single
wavenumber only) are not found.
These problems can be overcome by the addition of cubic terms 
to the amplitude equations.
In general, this is inconsistent, since quadratic and cubic terms can
only appear together if the amplitudes are $O(1)$, in which case terms
of all order should appear in the equations.
However, a consistent scaling can be obtained if the asymmetry in the
problem (and hence the quadratic term) is small; this is a commonly used
assumption \cite{Cox97,Hall95,Kub96}.
The appropriate scaling is that the coefficient of the quadratic term
is $O(\ep)$ (recall that $\ep$ is the scale of the difference between
the wavenumbers). The growth rates are $O(\ep^2)$ as before, and a
consistent balance between linear, quadratic and cubic terms is
obtained if the amplitude scaling is $A_i = O(\ep)$.
The scaled amplitude equations then become
\begin{eqnarray}
\dot A_1 &  = & \la_1 A_1 + A_2 A_3 - A_1( A_1^2 + \beta A_2^2 +
\beta A_3^2) \nonumber \\
\dot A_2 &  = & \la_2 A_2 + A_1 A_3 - A_2( A_2^2 + \beta A_3^2 +
\beta A_1^2)  \label{eq:ampcub} \\
\dot A_3 &  = & \la_3 A_3 + A_1 A_2 - A_3( A_3^2 + \beta A_1^2 +
\beta A_2^2) .\nonumber   
\end{eqnarray}
It can be assumed that the quadratic and
cubic terms are equal in each equation because the system
(\ref{eq:ampcub}) represents a small perturbation from the equations
for regular hexagons; any deviation appears at higher order.
By choosing an appropriate scaling for both time and the amplitudes,
the coefficients of the quadratic terms and the $A_i^3$  terms have
been set to unity.
The constant $\beta$ is  problem--dependent and cannot be
scaled out. For simplicity it is assumed that 
$\beta > 1$. This means that rolls are stable in the absence of the
quadratic term. This is indeed the case for both Rayleigh--Benard
convection and for the Swift--Hohenberg equation.

\begin{figure}
\centerline{
\epsfxsize0.95\hsize\epsffile{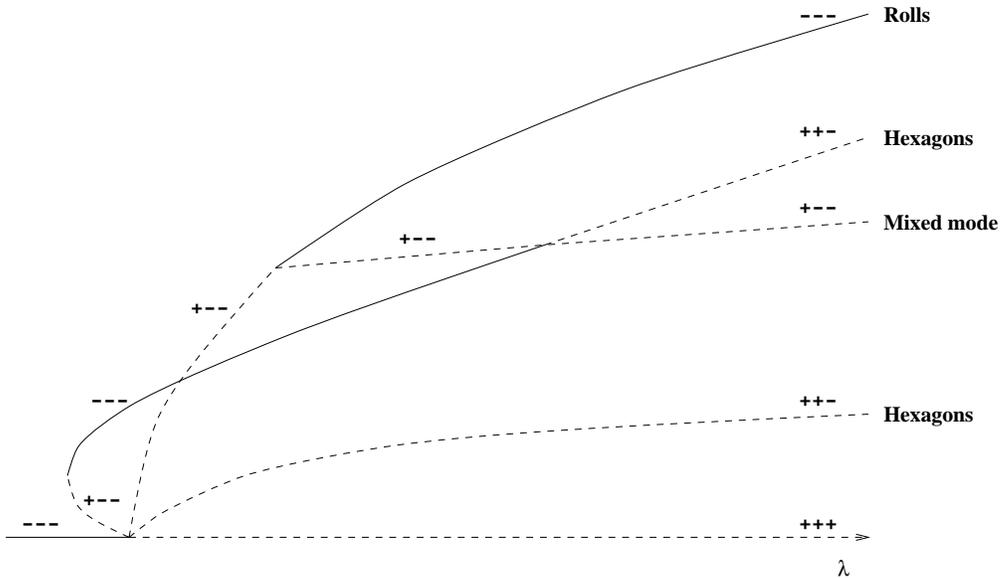}
}
\caption{Bifurcation diagram for regular hexagons. Solid lines
indicate stable solutions and dashed lines represent unstable
solutions. Pluses and minuses show the signs of the three eigenvalues
of each branch.\label{fig:d3bifc}}
\end{figure}

It is helpful to consider first the familiar case of regular hexagons 
for which $\la_1 = \la_2 = \la_3$ \cite{Gol84}. The bifurcation
diagram for this case is shown in Fig.~\ref{fig:d3bifc}.
Solutions in the form of rolls 
(e.g. $A_1 = \pm \sqrt {\la_1}$, $A_2 = A_3 = 0$) are stable for 
$\la_1 >  1/( \beta- 1 )^2$. Hexagons (with $A_1 = A_2 = A_3
\ne 0$) appear at a transcritical bifurcation from the trivial
solution where they are unstable, but gain stability though a
saddle--node bifurcation at $\la_1 = -1/(4 + 8 \beta) = \la_{\rm
SN}$ and are stable for 
$\la_{\rm SN} <  \la_1 < \la_{\rm TR} = (\beta + 2)/( \beta - 1)^2 $.
There are two regions of overlapping stable solutions and hence
hysteresis; for $\la_1$ small and negative both the trivial solution and
hexagons are stable, while for larger $\la_1$ both hexagons and rolls
are stable. The rolls and hexagons are linked via a branch of mixed
modes (e.g. $A_1 = A_2 \ne A_3$) which are always unstable.
The upper stability boundary of hexagons at $\la_1 = \la_{\rm TR}$ 
occurs at a transcritical
bifurcation with $D_3$ symmetry, where the hexagons and mixed modes
have a double zero eigenvalue; a centre manifold reduction near this
point yields the normal form 
\beq \dot x  =  \mu x + x^2 - y^2 , \qquad \dot y = \mu y - 2 x y 
\label{eq:d3trans}\eeq
with the symmetries of rotation through $2\pi/3$ and reflection $y \to - y$.
This system is discussed in more detail later.

The procedure for analysing (\ref{eq:ampcub}) is similar to that for 
(\ref{eq:amp}).
The behaviour depends on whether or not any of $\la_1$, $\la_2$ and $\la_3$
are equal, so these two cases are studied separately.
Since the equations (\ref{eq:ampcub}) have been considered by other
authors \cite{Mal94,Hall95}, the details of the calculations are not
given. Attention is focussed on points not described by the previous work.
The case of symmetric patterns with two equal wavenumbers is
described in section \ref{sec:D2c} and the asymmetric case with three
different wavenumbers is discussed in section \ref{sec:Z2c}. 

\subsection{Patterns with two equal wavenumbers}
\label{sec:D2c}

The fixed points of (\ref{eq:ampcub}) when $\la_1 = \la_2$ are as follows.\\
(i) $A_1 = A_2 = A_3 = 0$. This is stable if $\la_1 <0$ and $\la_3 < 0$, 
and undergoes stationary bifurcations at $\la_1 = 0$ and $\la_3 =  0$. \\
(ii) `$\la_1$-rolls' with $A_1 = \pm\sqrt{\la_1}$, $A_2 = A_3 = 0$; or
$A_2 = \pm\sqrt{\la_1}$, $A_1 = A_3 = 0$.
These rolls are appear at a supercritical pitchfork bifurcation
and are stable if $\la_1 > ( 1 - \la_3 +\la_3 \beta )/\beta(\beta-1)$. \\
(iii) `$\la_3$-rolls' with $A_1 = A_2 = 0$, $A_3 = \pm\sqrt{\la_3}$.
These rolls also bifurcate supercritically, are stable  if
$\la_1 <  \beta\la_3  - \sqrt{\la_3}$, 
and have bifurcations at 
$\la_1 = \beta\la_3 \pm\sqrt{\la_3}$. \\
(iv) A mixed mode which can be regarded as hexagons, for which  $A_1=\pm A_2$.
This solution  also branches from the
zero solution at a pitchfork bifurcation at $\la_1 = 0$,
which is supercritical if 
$1  + \beta + 1/\la_3 > 0$. At onset, $A_3$ is much smaller
than $A_1$ and $A_2$, so this  pattern has a rectangular appearance.
This solution can undergo saddle-node bifurcations, and branches from
the  $\la_3$-rolls at $\la_1 = \beta\la_3 \pm\sqrt{\la_3}$. \\
(v) A mixed mode in which all the amplitudes are different.
This solution branches from the $\la_1$-rolls at $\la_1 = ( 1 - \la_3
+\la_3 \beta )/\beta(\beta-1)$, and can never be stable.

\begin{figure}
\centerline{
\epsfxsize0.95\hsize\epsffile{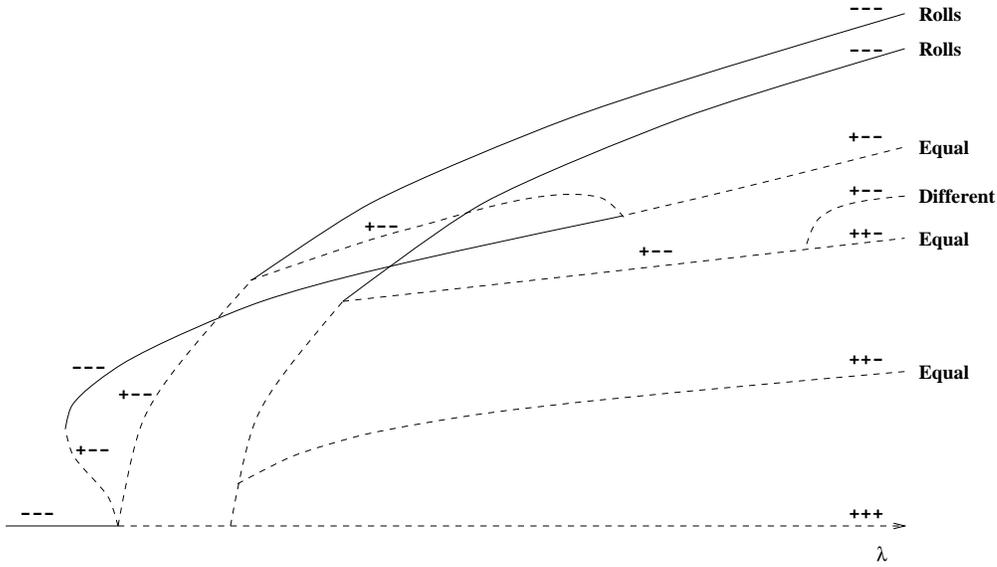}
}
\caption{Bifurcation diagram for  hexagons with mirror symmetry in the
case where the double mode bifurcates first ($\delta > 0$). 
The mixed modes are labelled according to whether all three amplitudes
are different, or two are equal.\label{fig:d2bifa}}
\end{figure}

\begin{figure}
\centerline{
\epsfxsize0.95\hsize\epsffile{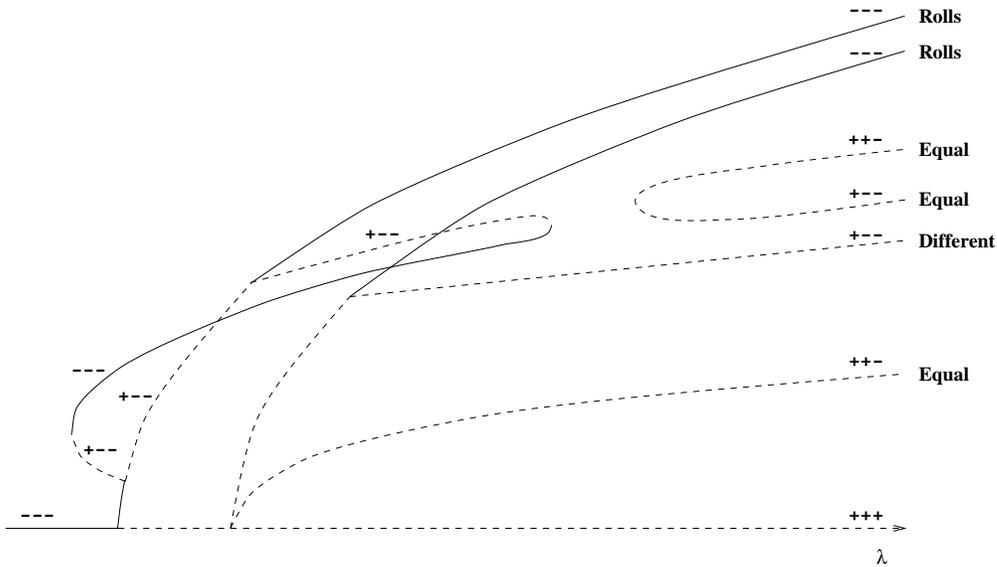}
}
\caption{Bifurcation diagram for  hexagons with mirror symmetry in the
case where the single mode bifurcates first ($\delta < 0$).\label{fig:d2bifb}}
\end{figure}

To draw the bifurcation diagram, the number of  parameters
can be reduced to one by supposing that $\beta$ and the box size $L$ are fixed
but $r$ is allowed to vary. This is equivalent to  writing 
$\la_1 = \la_3 +\delta$, with $\delta$ fixed. 
There are two possible cases, according to the sign of $\delta$,
and these are shown in Fig.~\ref{fig:d2bifa} and Fig.
\ref{fig:d2bifb}. In the bifurcation diagrams 
it is assumed that $\delta$ is small, so that the bifurcations from
the zero solution are close together. 

In the case $\delta > 0$ (Fig.~\ref{fig:d2bifa}) 
the double bifurcation from the origin
occurs first as $\la_1$ is increased. If $\delta < 1/(1+\beta)$, 
a mixed mode with $A_1=\pm A_2$
bifurcates subcritically from the origin and then gains stability in a
saddle-node bifurcation. For $\delta >  1/(1+\beta)$, the mixed mode
bifurcates supercritically.
This  mixed mode loses stability when it
bifurcates to the (unstable) solution with all amplitudes different.
Both roll solutions become stable at large $\la_1$. 

For $\delta <  0$ (Fig.~\ref{fig:d2bifb}), $\la_3$-rolls are stable
at small amplitude but lose stability at a pitchfork  bifurcation 
to the mixed mode with $A_1=\pm A_2$. This bifurcation is subcritical
for small values of $|\delta|$.  The mixed mode gains stability
at a saddle-node bifurcation but then loses stability at a second
saddle-node bifurcation and rejoins the roll branch, at which point
the rolls regain stability. For larger values of $|\delta|$ the region
of stable mixed modes decreases and for $\delta < -1/4(\beta-1)$ there
are no stable mixed modes, so that rolls are always stable.

\begin{figure}
\centerline{
\epsfxsize0.95\hsize\epsffile{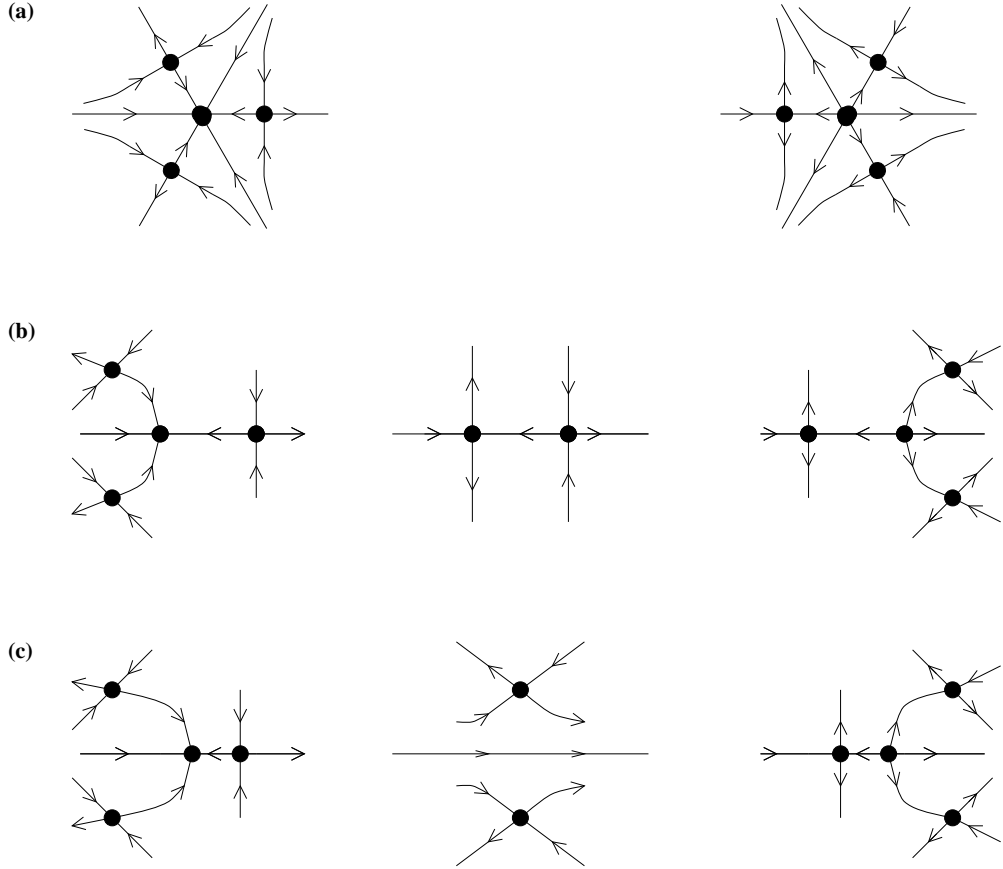}
}
\caption{The splitting or `unfolding' of the transcritical bifurcation
with broken $D_3$ symmetry (\ref{eq:d2trans}). Phase diagrams are shown 
with $\mu$ increasing from left to right, for the three cases:
(a) $\delta=0$: Three mixed modes cross through the hexagons.
(b) $\delta>0$: Two mixed modes with all amplitudes different
disappear at a pitchfork bifurcation with a mixed mode with two
amplitudes equal, then reappear at a pitchfork bifurcation with the
other mixed mode.  (c) $\delta<0$: The two mixed modes with  two
amplitudes equal disappear and then reappear at two saddle-node
bifurcations.\label{fig:d2trans}}
\end{figure}

It is of interest to note that these diagrams can be determined to a
considerable extent just by considering the splitting of the 
hexagonal case (Fig.~\ref{fig:d3bifc}) induced by the
symmetry-breaking. The single branch of rolls splits into two
branches, and both hexagon branches become mixed modes with $A_1=\pm A_2$. 
The mixed mode with $A_1=\pm A_2$ is unaltered, while that with 
$A_1=\pm A_3$ becomes a mixed mode with all amplitudes different. 
It follows that in both the cases $\delta >  0$ and $\delta <  0$, 
at large $\la_1$ there must be two stable roll solutions, three mixed
modes with $A_1=\pm A_2$ (two with a $++-$ stability and one with
$+--$), and one mixed mode with all amplitudes different. 
This approach can be used to construct Figs. \ref{fig:d2bifa} and
\ref{fig:d2bifb} except near $\la_1=0$, where the amplitudes are
small,  and near $\la_1 = \la_{\rm TR} = (\beta+2)/(\beta-1)^2$. 
Near this latter point, where in the case of regular hexagons there 
is a transcritical bifurcation (\ref{eq:d3trans}) with $D_3$ symmetry, 
the rotational symmetry is broken, but the reflection symmetry ($y \to
- y$) is retained. A centre manifold reduction  of (\ref{eq:ampcub})
near $\la_{\rm TR}$ for small $\delta$ leads to the normal form
\beq \dot x  =  \mu x + x^2 - y^2 - \delta, \qquad 
     \dot y  =  \mu y - 2 x y , 
\label{eq:d2trans}\eeq
where $\mu$ is proportional to $\la_1 - \la_{\rm TR}$ and
$x$ and $y$ are rescaled forms of $2 A_3 - A_1 - A_2$ and $A_1 - A_2$
respectively. 
In the case of regular hexagons ($\delta=0$) three mixed modes cross
through the hexagons as $\mu$ passes through zero and the phase
portraits are as shown in Fig.~\ref{fig:d2trans}(a).
For $\delta \ne 0$ the behaviour of (\ref{eq:d2trans}) depends on the
sign of $\delta$. 
For $\delta > 0$, two pitchfork bifurcations occur at 
$\mu = \pm 2 \sqrt{\delta/3}$ and the sequence of phase portraits 
is as shown in Fig.~\ref{fig:d2trans}(b). 
For $\delta < 0$, two saddle-node bifurcations occur at $\mu = \pm 2
\sqrt{-\delta}$; the phase portraits are shown in Fig.
\ref{fig:d2trans}(c). 
This analysis enables  the correct connections to be made to the
various branches in the Figs. (\ref{fig:d2bifa}) and (\ref{fig:d2bifb}).

\subsection{Patterns with three different wavenumbers}
\label{sec:Z2c}
 
In the case where the pattern involves three different wavenumbers, 
$\la_1$, $\la_2$ and $\la_3$ are all different. The fixed points of  
(\ref{eq:ampcub}) are: \\
(i) $A_1 = A_2 = A_3 = 0$. This is stable if $\la_1 <0$, $\la_2 <0$
and $\la_3 < 0$,  and undergoes stationary bifurcations at $\la_1 =
0$, $\la_2 = 0$ and $\la_3 =  0$. \\ 
(ii) Three branches of rolls, for example $A_1 = \pm\sqrt{\la_1}$,
$A_2 = A_3 = 0$. 
These are stable if $\la_1 + \la_2 < 2\beta \la_1$ and
$(\la_2 - \beta \la_1)(\la_3 - \beta \la_1) > \la_1$. \\
(iii) A mixed mode in which the three amplitudes are all different,
which bifurcates from the roll solutions.

With a fixed box size, we can set $\la_1 = \la_2 + \alpha = \la_3 +
\delta$ with $\alpha$ and $\delta$ fixed. Since the three modes are
equivalent, it can be assumed without loss of generality that 
$\delta > \alpha > 0$, so that the rolls bifurcate in the order 
$A_1$, $A_2$, $A_3$ as $\la_1$ increases.
The following results can then be obtained for the roll solutions:
The first branch of rolls to bifurcate is stable at onset. 
For small $\alpha$ and $\delta$ these rolls lose stability and then
regain stability as $\la_1$ increases; for larger  $\alpha$ and
$\delta$  these rolls are always stable.
The second branch of rolls to bifurcate is unstable at onset and
becomes stable at a bifurcation to a mixed mode.
The third branch of rolls is also unstable at onset, has two
bifurcations to mixed modes and is then stable.

\begin{figure}
\centerline{
\epsfxsize0.95\hsize\epsffile{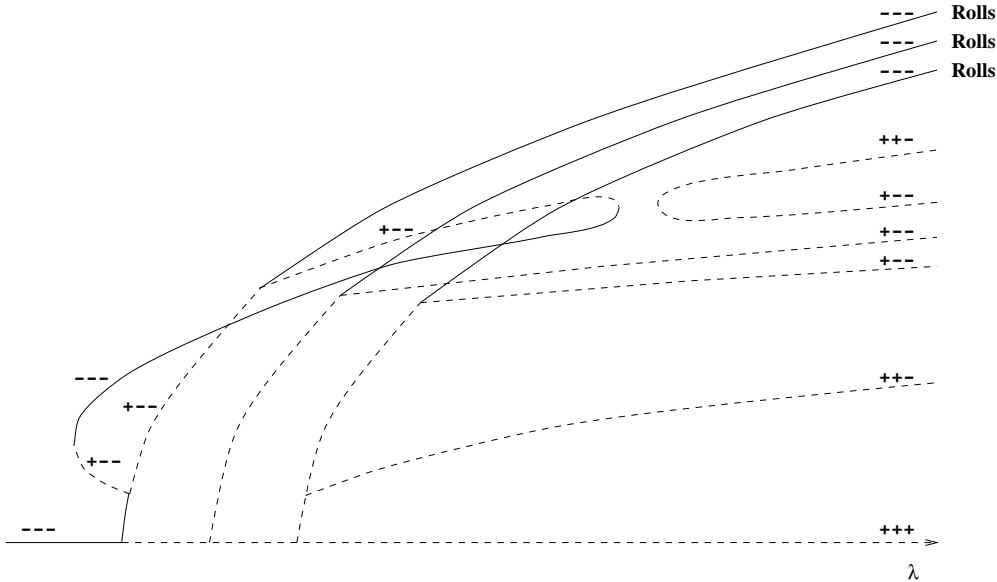}
}
\caption{Bifurcation diagram for hexagons without mirror
symmetry.\label{fig:z2bif}  }
\end{figure}

The bifurcation diagram for small $\alpha$ and $\delta$ is shown in
Fig.~\ref{fig:z2bif}. There are up to five distinct branches of
mixed modes. Three of these correspond to the  mixed modes in
Fig.~\ref{fig:d3bifc} and the other two correspond to the hexagons.
Near the transcritical bifurcation at $\la_1 = \la_{\rm TR}$ a
centre manifold reduction yields a second-order system 
analogous to (\ref{eq:d2trans}):
\beq \dot x  =  \mu x + x^2 - y^2 - \gamma_x, \qquad 
     \dot y  =  \mu y - 2 x y - \gamma_y. 
\label{eq:z2trans}\eeq
Here, both the rotation and reflection symmetries have been broken.
The constants $\gamma_x$ and $\gamma_y$ are related to $\alpha$ and
$\delta$  but their value is not important. 
The behaviour of (\ref{eq:z2trans}) is similar to that shown in  Fig. 
\ref{fig:d2trans}(c) (but without the mirror symmetry), 
so that two of the mixed modes undergo a pair of saddle-node
bifurcations, while two undergo no bifurcations, as 
shown in Fig.~\ref{fig:z2bif}.

\section{Numerical simulations of the asymmetric Swift--Hohenberg equation}
\label{sec:num}

To illustrate the results of the preceding work, this section
describes numerical simulations of the Swift--Hohenberg equation
\cite{Swift77} modified by the addition of a quadratic term:
\beq
\prt w t  = r w - (1+\nabla^2)^2 w + s w^2 - w^3 . \label{eq:sh}
\eeq
This equation is probably the simplest with the required properties of
a preferred wavenumber ($k=1$) and an asymmetry between $u$ and $-u$,
and (\ref{eq:sh}) or variations of it are commonly used as models for
pattern formation (e.g. \cite{Kub96}).
A Fourier mode with wavenumber $k$ has growth rate
$r-(1-k^2)^2$.
By substituting three Fourier modes of the form (\ref{eq:expand}) into
(\ref{eq:sh}), the amplitude equations (\ref{eq:ampcub}) are obtained
with $\beta=2$ and $\lambda_i = 3(r-(1-k_i^2)^2)/4 s^2$.

A Fourier spectral method was used to solve (\ref{eq:sh}) with
periodic boundary conditions in the region $0 < x, y < L$.
In Fourier space the linear parts of  (\ref{eq:sh}) can be solved
exactly since the Fourier modes decouple. The nonlinear terms were
integrated using the second-order Adams--Bashforth method.
The initial condition chosen was a small-amplitude random 
perturbation to the zero solution.

\begin{figure}
\hskip1.3truein (a)\hskip2.5truein (b)
\centerline{
\epsfxsize0.48\hsize\epsffile{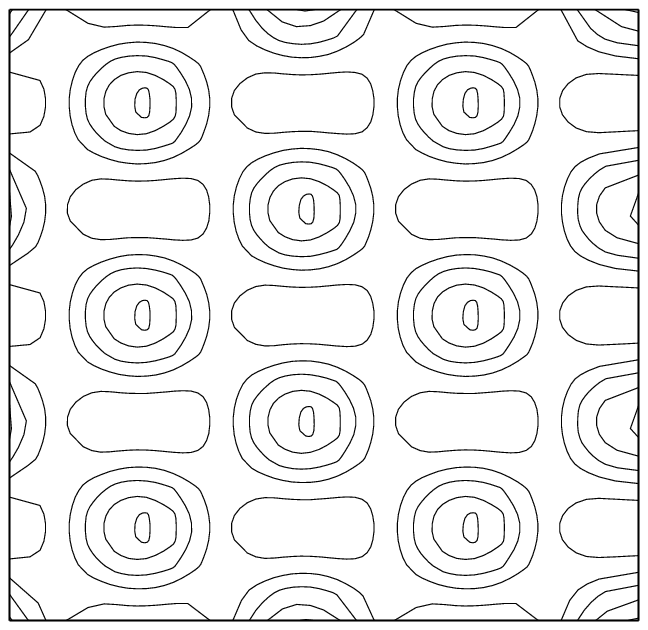}\epsfxsize0.48\hsize\epsffile{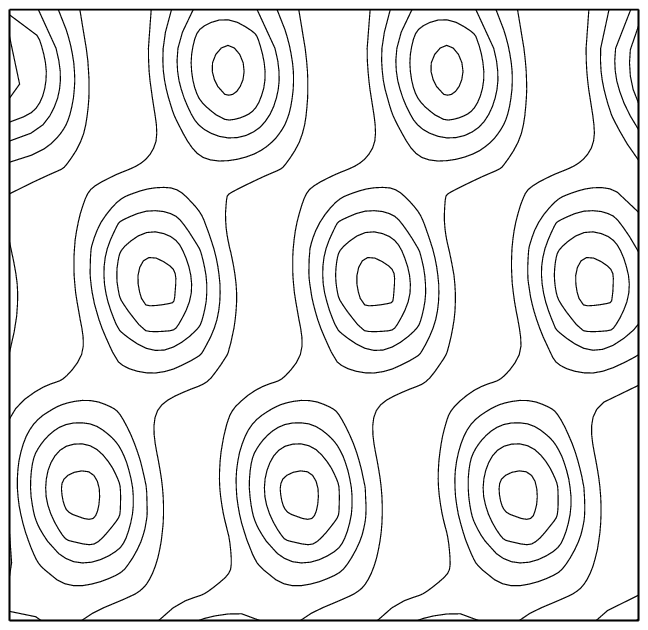}
}
\caption{Numerical solutions to  (\ref{eq:sh}) with $s=0.25$.
(a) $r=0.03$, $L=23$, showing 12 irregular hexagons.
(b) $r=0.05$, $L=21.25$, showing 9 irregular hexagons.\label{fig:h129}}
\end{figure}

For the case of hexagons with reflection symmetry, the pattern with 12
hexagons with wave integers $(4,0)$, $(-2,3)$ and $(-2,-3)$ was
studied. In this case, there are two different possible bifurcation
diagrams (Figs. \ref{fig:d2bifa} and \ref{fig:d2bifb}) according to
whether the double mode or the single mode bifurcates first.
For $L=23$, the $(-2,3)$ and $(-2,-3)$ modes bifurcate first and therefore
the appropriate bifurcation diagram is Fig.~\ref{fig:d2bifa}.
However for these parameters, $\delta = 0.445$, so from the 
analysis of section \ref{sec:D2c} we expect the bifurcation 
to irregular hexagons to be supercritical.
The numerical results show 12 hexagons for $0.002 \le r \le 0.03$ and 
pure $(3,2)$ rolls for $r \ge 0.05$, with both solutions stable for
$r=0.035$ and $r=0.04$. This is consistent with Fig.~\ref{fig:d2bifa}.
The solution for $r=0.03$ is shown in Fig.~\ref{fig:h129}(a).

For $L=24.5$, the $(4,0)$ mode bifurcates first and 
the appropriate bifurcation diagram is Fig.~\ref{fig:d2bifb}.
Rolls with wave integers  $(4,0)$ were found for 
$0< r \le 0.01$ and $r \ge 0.025$, and a hexagonal pattern including
the $(-2,3)$ and $(-2,-3)$ modes was found for $0.015 \le r \le 0.02$.

The solution with 9 hexagons, involving the wave integers 
$(3,2)$, $(-3,1)$ and $(0,-3)$ was chosen to study the case 
without reflection symmetry. This solution was found when the box size $L$ was
21.25, and at this value of $L$ the $(-3,1)$ mode bifurcates first.
A solution with $9$ hexagons was found for $0.02 \le r \le 0.055$ 
and rolls with wave integers $(3,1)$ were found for $r \ge 0.06$.
For $r=0.057$ both solutions were found to be stable.
Referring to the bifurcation diagram (Fig.~\ref{fig:d2bifb}),
rolls are also expected for small $r$; in fact however a hexagonal
solution with 8 hexagons was found for $r \le 0.02$.
Interactions between hexagonal patterns with different numbers of
hexagons are of course beyond the scope of the analysis of section
\ref{sec:dyn}. 
The solution for $r = 0.05$ is shown in Fig.~\ref{fig:h129}(b).

\begin{figure}
\centerline{
\epsfxsize0.98\hsize\epsffile{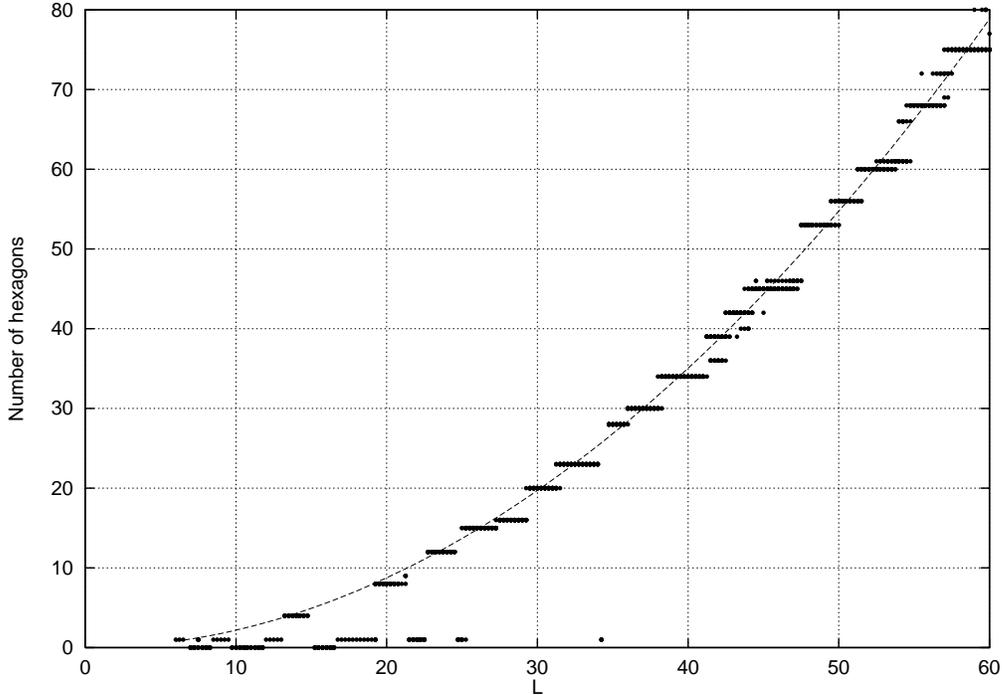}
}
\caption{Number of hexagons $N$ against box size $L$ for the equation
(\ref{eq:sh}) with $r=0.02$ and $s=0.25$. $N=0$ indicates
that the zero solution is stable and $N=1$ indicates that rolls
are stable.  The dashed line is the curve 
$N = \protect\sqrt{3} L^2 / 8 \pi^2$, obtained by dividing the area of the box
by the area of a regular hexagon. 
\label{fig:nhex}}
\end{figure}

One question of interest which is difficult to approach analytically 
is which of the many possible hexagonal solutions listed in table
\ref{tab:hex} are observed as the box size $L$ is increased.
To investigate this question a sequence of runs was carried out 
with $r=0.02$ and  $s=0.25$. For each value of $L$, the computation 
was continued until a stationary solution was obtained.
By looking at the dominant Fourier modes the solution was  then
identified as either rolls or hexagons and the number of hexagons was
determined using (\ref{eq:nhex}). Since there is the possibility of
multiple stable states, the computation was repeated ten times for
each value of $L$, with a different small-amplitude random initial
condition in each case. Values of $L$ between 6 and 60 were used, with
a step of 0.25 in $L$. The results are summarised in Fig.
\ref{fig:nhex}, which plots the number of hexagons against the box
size. Where the number of hexagons is zero this indicates 
the zero solution,  and a solution in the form of rolls is shown as 
one hexagon.  
Note that there are regions in which both rolls and hexagons
are stable, and also regions in which two or even three 
different types of hexagons are stable.

Of the hexagonal solutions  in table \ref{tab:hex},
those with 4, 8, 9, 12, 15, 16, 20, 23 hexagons were found.
Note that, as might be expected, these are the solutions with 
low values of $H$, i.e.\ those that are closest to regular hexagons. 
There  appears to be a preference for the symmetric hexagons.
However this is no longer true at large values of the box size,  where
there are roughly as many asymmetric patterns as symmetric ones.

\section{Conclusions}

The main results of this paper are as follows.
For numerical experiments with periodic boundary conditions, 
or laboratory experiments with Neumann  boundary conditions, 
a perfectly regular hexagonal pattern is not permitted in a square box
or any box with rational aspect ratio. 
Instead, patterns of irregular or `non-equilateral' hexagons 
are observed. These patterns are composed primarily of three
wavenumbers, and occur in two types. If two of the wavenumbers are
equal then the resulting pattern  has a mirror symmetry, but if all
the wavenumbers are different it does not. 
There is a simple relationship (\ref{eq:nhex}) between the number of
hexagonal cells in the pattern and  the three wavenumbers.

The nonlinear dynamics of these patterns is controlled by 
the amplitude equations (\ref{eq:ampcub}). 
The results of the analysis of these equations are summarised in the
bifurcation diagrams of Figs.~\ref{fig:d2bifa}, \ref{fig:d2bifb} and
\ref{fig:z2bif}. 
These provide a more complete picture of the dynamics than has been
given by previous studies of the equations
\cite{Cox97,Hall95,Hoyle96,Mal94}. 
A useful technique for clarifying the bifurcations that occur is the
centre manifold reduction near the transcritical bifurcation with
$D_3$ symmetry.
This reduces the algebra of the analysis considerably, enabling firm
conclusions to be drawn regarding the connections between different
solution branches.

The analysis is closely related to that of Proctor and Matthews
\cite{Proc96} who investigated the interaction of modes with wave integers
$(0,1)$, $(1,0)$ and $(-1,-1)$, which give a pattern with the topology
of hexagons (isolated upflows and connected downflows) but with square
symmetry. Some of the results are similar, for example the fact that
the quadratic resonances can lead to subcritical behaviour 
and a preference for three-dimensional patterns over two-dimensional rolls.

This work was motivated by numerical simulations of compressible
magnetoconvection \cite{Weiss96}, in which hexagonal patterns with six
rising plumes were found. 
The results obtained suggest that for very small Rayleigh numbers,
rolls should be stable, and that for larger Rayleigh numbers rolls
should again become stable.
However it must be borne in mind that the equations (\ref{eq:ampcub})
are only valid if the up--down asymmetry (the departure from the
Boussinesq approximation in the case of convection experiments) is
small, and this is generally not the case. 
There is much more complicated dynamics in these numerical experiments
and further analytical work is required to understand and interpret
them fully.

\section*{Acknowledgements}

I am grateful to the University of Nottingham, the Royal Society and
the Nuffield Foundation for financial support. This work has benefited
from discussions with S.~Cox, M.~Proctor, A.~Rucklidge and N.~Weiss.

\end{document}